Nonlinear current-voltage behavior and electrically driven phase transition in charge frustrated LuFe$_2$O$_4$

L.J. Zeng[1], H. X. Yang [1], Y. Zhang[1], H. F. Tian[2], C. Ma[1], Y. B.Qin[1], Y. G. Zhao[2] & J. Q. Li [1]*

[1] Beijing National Laboratory for Condensed Matter Physics, Institute of Physics, Chinese Academy of Sciences, Beijing 100190, P. R. China.

[2] Department of Physics, Tsinghua University, Beijing 100084, P. R. China

Electric transport measurements of the charge frustrated LuFe$_2$O$_4$, in which the charge ordering (CO) and electronic ferroelectricity are found, reveal strong nonlinear electric conduction upon application of electrical field in both single crystalline and polycrystalline samples. The threshold electric fields ($E_t$) in single crystalline LuFe$_2$O$_4$ are estimated respectively to be about 60V/cm and 10V/cm with E parallel and perpendicular to the c-axis direction. Experimental measurements also demonstrate that the I-V nonlinearity increases quickly with lowering temperature. Furthermore, our in-situ TEM investigations evidently reveal that the nonlinear I-V behavior is intrinsically in correlation with a current driven charge ordering insulator-metal transition, and the applied electrical field triggers a visible CO collapse recognizable as the fading of satellite spots of the CO modulations.

The charge-frustrated $RFe_2O_4$ system (R=Y, Er, Yb, Tm and Lu) has been extensively studied due to its rich variety of interesting physical properties and immense potential for technical applications, such as the spin/charge ordering [1], the multiferroic and the large magnetodielectric effects [2]. It is noted that the frustrated spin/charge system shows not only complex magnetic phase but also drive some new classes of physical phenomena [3, 4]. $LuFe_2O_4$ has a layered rhombohedral structure with the space group of R-3m, and lattice parameters of a=3.444Å and c=25.259 Å. The average valence of Fe ions is 2.5, which implies that the $Fe^{2+}$ and $Fe^{3+}$ occupy the equivalent Fe sites on the hexagonal net plane with an equal probability. In previous publications, a number of theoretical and experimental investigations on this kind of materials were focused on the strong coupling between spin and charge, and the room-temperature magnetodielectric response [2]. The charge ordering (CO) transtion at $T_c$~510K in $LuFe_2O_4$ is considered to be the origin of electric polarization for observed ferroelectricity [5]. Our TEM observations demonstrated that the ground state of the charge ordered phase can be fundamentally characterized as a real space arrangement of $Fe^{2+}$, $Fe^{2.5+}$ and $Fe^{3+}$, and the charge concentration has a nonsinusoidal charge density wave (CDW) characteristic [6]. Actually, the charge/spin frustrations in the ground state also play a critical role for the understanding of the magnetic order and the electronic ferroelectricity. Recently, we have performed a systematical study on the transport properties of $LuFe_2O_4$ materials upon application of electrical fields, notable nonlinear behaviors are commonly observed in a large temperature range on both polycrystalline and single crystalline samples. The nonlinear coefficients of single crystal can be estimated to be 200 and 30 with currents flowing parallel and perpendicular to c-axis direction, respectively. These data are much larger than that obtained from ZnO which is commonly used as metal-oxide varistors and ZnO-based ceramic devices with highly nonlinear current-voltage characteristics. Hence, an extensive study on the specific properties of $LuFe_2O_4$ materials may lead to novel technological applications for the developments of efficient switching devices and tunable surge protection varistors. Furthermore, in-situ TEM measurements with an applied voltage performed on a specific designed TEM holder reveal that this strong nonlinear behavior appears in association with a remarkable charge order-disorder transition. This electrically driven charge order to metallic phase transition shows certain similarity in comparable with what observed in magnetite $Fe_3O_4$ nano-structures [7] and a number of manganites [8, 9]. As a result, our study will shed light on understanding the unusual transport properties in the charge ordered stripe and checkerboard phases in strongly correlated systems, the fundamental properties of nonlinear conduction and the CO melting in the electrically driving non-equilibrium transition.

Both the single- and poly-crystalline $LuFe_2O_4$ samples were used in present study. The polycrystalline materials were synthesized by a conventional solid-state reaction under a controlled oxygen partial pressure atmosphere using a $CO_2$-$H_2$ mixture as reported in [10]. In order to obtain the $LuFe_2O_4$ single crystals, the $LuFe_2O_4$ powder was heated to 1620℃ in a platinum crucible, and then the melted solid was cooled to 900 °C at a rate of 1°C/min. The whole process is carried out under a $CO_2$-$H_2$ mixture of $CO_2/H_2$=2. The typical size of the grown single crystals was $1\times2\times0.5$ mm$^3$. The current-voltage (I-V) curves of the samples were measured by a two-probe method with a Keithley 2400 source meter. In order to protect the samples, a compliance current of 200mA was used. Silver contact pads with a radius of ~ 1mm were deposited on top of the samples by magnetron sputtering using a shadow mask. The resistivity of the sample was calculated from R=V/I. Transmission electron microscopy (TEM) investigations were performed on a Tecnai F20 (200kV) electron microscope equipped with a specific double tilt TEM holder on which in-situ observations can be carried out upon an applied voltage.

The fundamental I-V characteristics of $LuFe_2O_4$ were firstly measured on a number of polycrystalline materials that show up remarkable multiferroic in the charge ordered state as discussed in previous publications [1,5,6]. The current density (J) and electric field (E) in measurements were also computed numerically for each sample.

Fig. 1a shows the plot of current density against electric field obtained from a well-characterized polycrystalline sample at room temperature, demonstrating the presence of the evident nonlinear J-E characteristics in the $LuFe_2O_4$ materials. It is known that the grain boundaries in the polycrystalline samples could produce variety kinds of potential barriers that could give rise to the additional effects in the measured resistivity. For instance, measurements of transport properties of $CaCu_3Ti_4O_{12}$ demonstrated that the grain boundaries in the polycrystalline samples could serve as obstacles to the current flow through the specimens and yield clear nonlinear I-V behaviors in the experimental data [11]. In order to understand the essential I-V characteristics of $LuFe_2O_4$ crystals, we therefore performed an extensive investigation on the well-characterized single crystalline samples.

Fig. 1b and 1c show the experimental J-E curves obtained from a single crystalline $LuFe_2O_4$ sample for the current flowing respectively parallel (E∥c) and perpendicular (E∥ ab plane) to the c-axis direction. These plots clearly demonstrate the presence of essential non-ohmic J-E characteristics both along the c-axis and within the ab plane in a $LuFe_2O_4$ crystal. Moreover, the strongly anisotropic electric transport properties in this layered system can be easily recognizable from the remarkable distinction of the data shown in Fig. 1b and 1c. Careful analysis reveals that the nonlinear behavior in each measurement can be well demonstrated by a clear anomaly at a threshold electric field ($E_t$) above which the current begins to abruptly increase. The $E_t$ in the single crystalline sample with E parallel and perpendicular to c axis are estimated to be about 60V/cm and 10V/cm, respectively. These data are far smaller than a typical breakdown field (~ $10^7$ V/cm) in a band insulator [12] and comparable with what observed in the charge ordered manganites [8,9].

The nonlinear current-voltage characteristic is often empirically described by the power-law relation, $I = kV^\alpha$, where k is a constant, $\alpha = d(\log I)/d(\log V)$ is the nonlinear coefficient. According to our experimental data, the nonlinear coefficients of the single crystal can be estimated to be about 200 and 30 for I∥c and I⊥c direction when measured in the current range 6-100mA, respectively. The log scale J-E curve is shown in the inset of each figure, clearly illustrating the nonlinear behaviors in both single crystalline and polycrystalline materials.

The observations of highly nonlinear characteristics in single crystals indicate that the nonlinear I-V behavior is an essential characteristic of the charge frustrated $LuFe_2O_4$ materials. Actually, according to the data shown in fig.1a , we can estimate the nonlinear coefficient for polycrystalline to be about 60 when measured in the current range 6-100mA, this result can be fundamentally understood as an average value for the randomly orientated single crystalline grains in a bulk material. This fact suggests that the grain boundaries in the $LuFe_2O_4$ materials may have a good conductivity in sharp contrast with that observed in the polycrystalline ZnO and $CaCu_3Ti_4O_{12}$ [13,11], in which the grain boundaries are considered to be responsible for the unusual I-V nonlinearity.

In order to understand the stability and recoverability of nonlinear behaviors in present system, we measured the I-V characteristics of the specimens repeatedly up to the upturn region on a few samples, the experimental results clearly show that the similar nonlinear behaviors can be cyclically obtained in each run, and, moreover, notable hysteresis in the I-V curves are often obtained as discussed in following context.  Fig. 2 shows three represented J-E plots of twelve measurements for a single crystalline sample with I∥c, and its inset illustrates the data at logarithmic scale. It is clearly recognizable that the remarkable nonlinearity is completely recoverable when the high voltages are removed.

The $LuFe_2O_4$ materials undergo a number of notable phase transitions in a large temperature range, such as the 3-dimensional CO and spontaneous electric polarization at $T_p$=350K, and a ferrimagnetic transition at $T_n$ =230K. Hence, we further carried out our investigations on the I-V nonlinearity at certain low temperatures.  Fig. 3 shows the current-voltage characteristics on a polycrystalline sample at temperatures of 300K, 250K, and 200K, respectively, demonstrating the presence of the strong nonlinear J-E relationships at the low temperature range. The most strike feature revealed in these data is that the $E_t$ increases markedly with lowering temperature: it is estimated to be about 50V/cm at 300K, 150V/cm at 250K, and 450V/cm at 200K. This fact suggests that the well-defined multiferroic state bellow $T_n$=230K is extremely favorable for sharp conductance switching in high electric field. The nonlinear coefficient $\alpha$ rises from $\alpha$ = 60 at 300K up to 400 at 200K. Fig. 3b illustrates a J-E hysteresis loop at 300K, showing evidently hysteretic feature, similar results were also obtained in the measurements on single crystalline samples.

In order to reveal the correlation between the nonlinear I-V behavior and the charge ordering observed in $LuFe_2O_4$, we performed an extensive in-situ TEM study by using a special designed TEM holder on which the microstructure properties and subsequent resistivity can be simultaneously recorded under an applied electric voltage. Figure 4a shows a series of electron diffraction patterns obtained under the applied voltages. The corresponding I-V data obtained simultaneously on the sample is shown in fig. 4b. Fig. 4a1 represents a typical [1-10] zone-axis diffraction pattern of $LuFe_2O_4$，showing the presence of superstructure reflections from two modulations (q1 and q2) as discussed in our previous publications [6, 10]. These structural modulations can be well interpreted by the $Fe^{3+}$, $Fe^{2.5+}$ and $Fe^{2+}$ ionic ordering which plays an important role for the multiferroic discovered in present materials.

The most salient feature revealed in the TEM observation is the speedy disappearance of the satellite spots as the applied voltage over the threshold voltage ($V_t$ = 28V) and then these spots reoccur again as the applied voltage removed. This fact directly suggests the occurrence of an electrically driven phase transition from the CO to a charge disordered state at $V_t$. This feature is qualitatively in agreement with the I-V measurements mentioned in above context. In combination with the nonlinear I-V data associated with the present TEM observations, it is apparently noted that the satellite spots become speedily faint as soon as the current I in the sample is detected and then disappear when the current exceeds a critical value (fig 4b). In general, the q2 modulation spots fade firstly away and，soon afterwards, the q1 modulation spots disappear in one or two minutes with the electric voltage being held at 28V. Moreover, as the applied voltage removed, q1 modulation spots reoccur immediately (fig 4a5), and then the q2 modulation spots reoccur in a few minutes (fig 4a6). Similar phenomena were found in all the grains we investigated.

The CO phenomenon and electronic ferroelectricity in $LuFe_2O_4$ are fundamentally in correlation with the strong on-site Coulomb repulsion of the 3d electrons in this frustration system [1], our in-situ TEM observations directly reveal that the applied electrical bias could efficiently drive the localized electrons out of equilibrium and result in an insulator to metal phase transition accompanying with a remarkable CO melting at around $V_t$. On the other hand, the current induced insulator–metal transition and the I-V nonlinearity were also discussed in other strongly correlated materials, such as the charge ordered $La_{0.5}Ca_{0.5}MnO_3$ [14], the organic molecular Mott insulator K-TCNQ [15] and the magnetite $Fe_3O_4$ [7], nevertheless, different models were proposed according to the specific experimental results in each system, and transition mechanisms remain highly controversial. For instance, resistance measurements on $La_{0.5}Ca_{0.5}MnO_3$ revealed dramatic hysteresis effects and broadband noise properties [14]. S. Cox et al interpreted this resistance behavior as a collective transport of a charge density wave. Chen et al has performed a series of measurements on the $La_{2/3}Ca_{1/3}MnO_3$ and $Fe_3O_4$ [16], they interpret the observed I-V nonlinearity in the regime of strong Joule heating. According to our careful TEM measurements in $LuFe_2O_4$, the CO state collapses immediately in association with an insulator to metallic phase transition, no visible collective movements of charge stripes is observed as proposed by Cox. Moreover, although current-heating effects should be taken into account in the data analysis, we have performed a number of measurements in which the sample temperature is well controlled at low temperatures as illustrated in figure 3, therefore, the Joule heating should not yield notable changes in addition to the essential nonlinearity.

In conclusion, we have demonstrated strong nonlinear I-V behavior in $LuFe_2O_4$ in both single crystals and polycrystalline materials, careful analysis shows that grain boundaries in polycrystalline samples do not play important role for the appearance of the nonlinear characteristic. The strong anisotropic feature origins from the layered structural feature is also observed in the I-V measurements, with the threshold electric fields being estimated to be about 60V/cm and 10V/cm with E parallel and perpendicular to c axis of the $LuFe_2O_4$ crystal. Experimental measurements show that the $E_t$ and nonlinear coefficients increase quickly with lowering temperature. Furthermore, our in-situ TEM observations evidently reveal that the nonlinear I-V behavior accompanying with a current driven charge ordered insulator-metal transition, CO collapse, in sharp contrast with the collective transport model, are directly documented by the disappearance of superlattice reflections. We anticipate that the observations of the large nonlinear coefficient, the conduction switching and CO melting will shed light on the physical mechanism of dynamic features of stripe and checkerboard phases linked to many exotic phenomena such as high-Tc superconductivity and colossal magnetioresistance. The large nonlinear coefficient and the conduction switching of $LuFe_2O_4$ could lead to certain novel technological applications in the developments of efficient switching devices and tunable surge protection varistors.

Acknowledgements

This work is supported by the National Science Foundation of China, the Knowledge Innovation Project of the Chinese Academy of Sciences, and the 973 project of the Ministry of Science and Technology of China.

Figure captions:

Fig. 1. Current density versus electric field (J-E) curves of LuFe$_2$O$_4$ measured at room temperature: (a) J-E plot of polycrystalline sample; (b) I-V characteristics of single crystalline LuFe$_2$O$_4$ with the applied electric field parallel to c axis; (c) J-E plot of single crystal with E perpendicular to c axis. The arrows denote the threshold electric fields for nonlinear conduction. In the insets of (a), (b), and (c), the log-log plots numerically gotten from the corresponding J-E curves are also shown. The nonlinear coefficient obtained from polycrystal is 60, while it is 30 and 200 with E parallel and perpendicular to c axis, respectively, when measured in the current range 6-100mA.

Fig. 2. Current density-electric field characteristics of single crystals with E parallel to c axis. The log scale plot is also shown in the inset. These curves show almost the same nonlinear character in repeated measurements.

Fig. 3. (a) Nonlinear current-voltage (I-V) characteristics of LuFe$_2$O$_4$ obtained at 300K (squares), 250K (circles), and 200K (triangles). (b) I-V hysteresis curve measured at room temperature. Arrows indicate the direction of hysteresis.

Fig. 4. Results of in-situ TEM measurement: (a) electron diffraction patterns with various bias voltages and currents, two different superlattice reflections are denoted as q1 and q2. The disappearance and recurrence of the superlattice reflections are clearly shown in a1-a6. (b) The intensity of superlattice reflection as a function of bias voltage, also shown is the I-V curve simultaneously obtained in the TEM measurement.

Fig 1

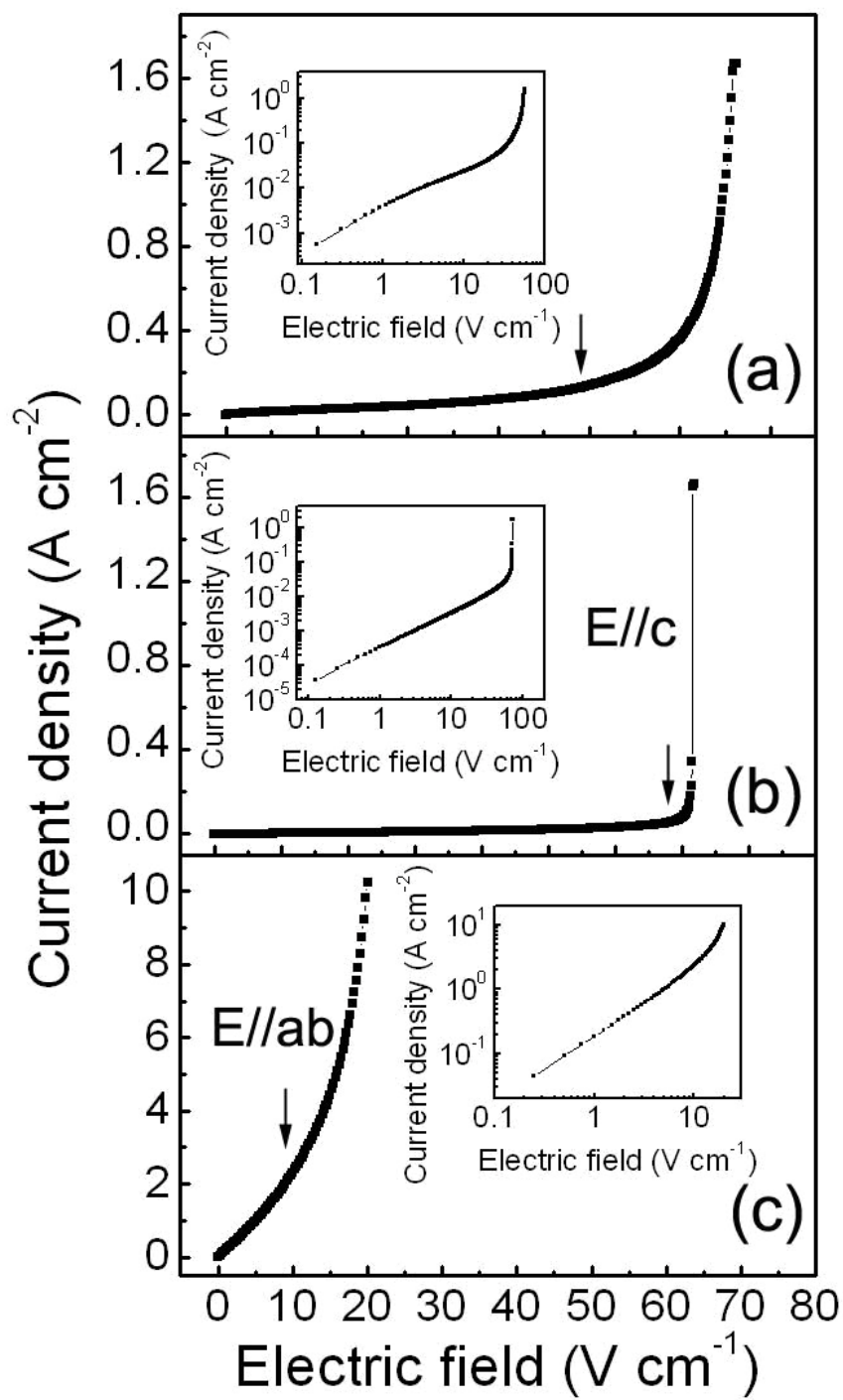

Fig 2

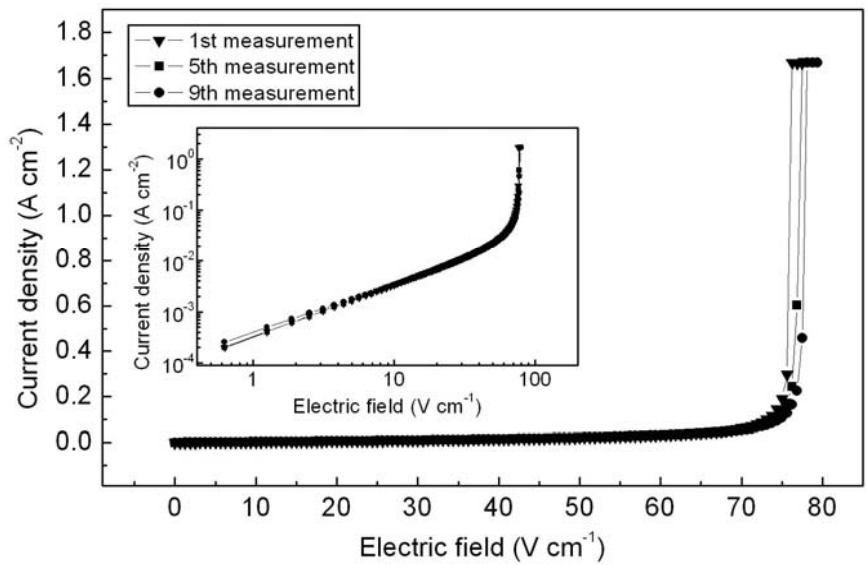

Fig. 3.

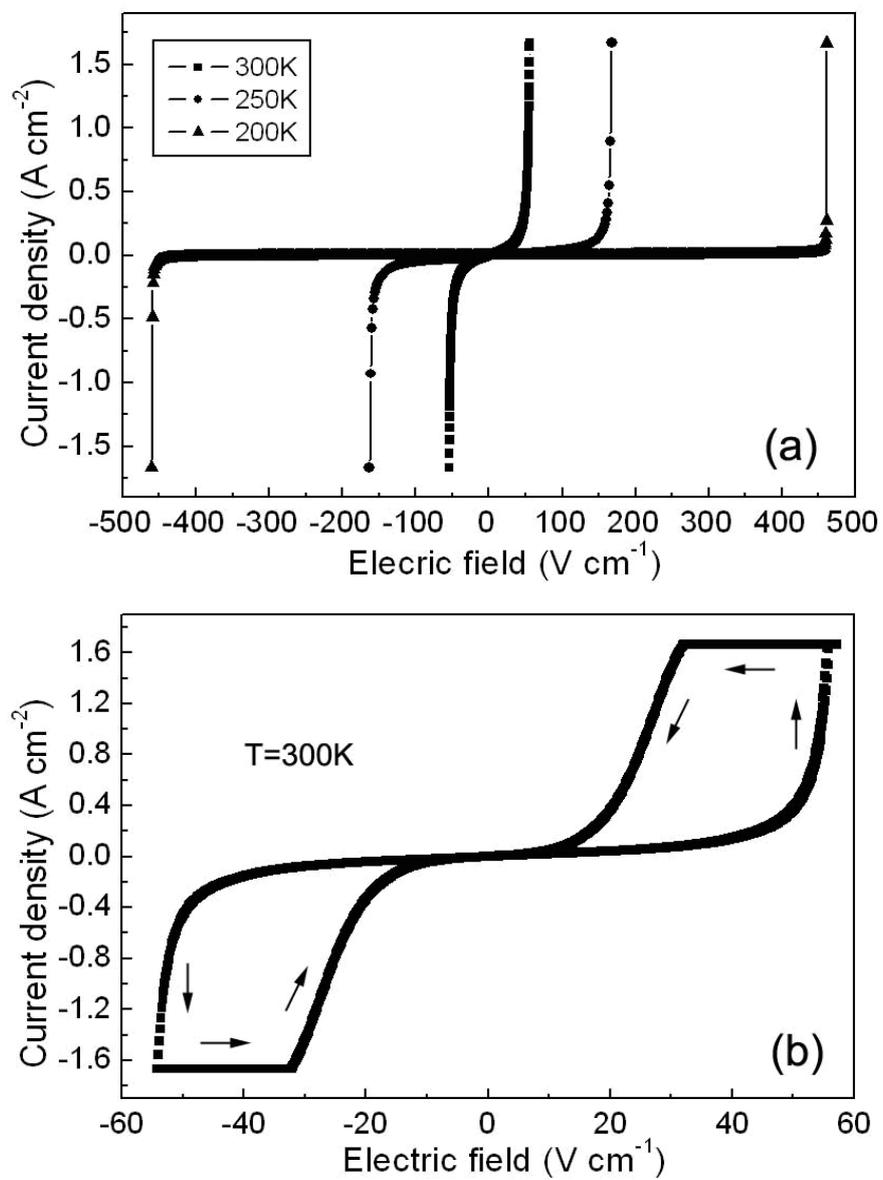

Fig. 4.

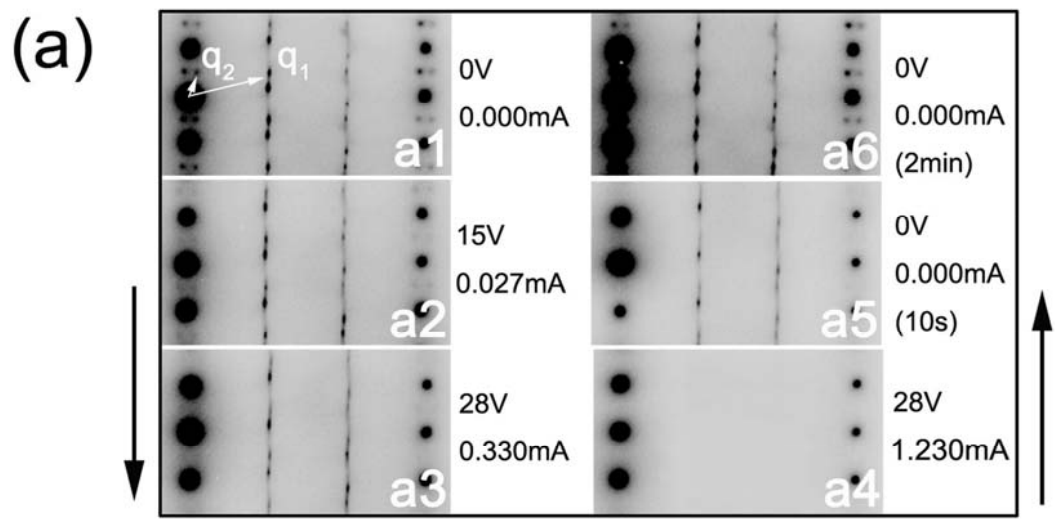

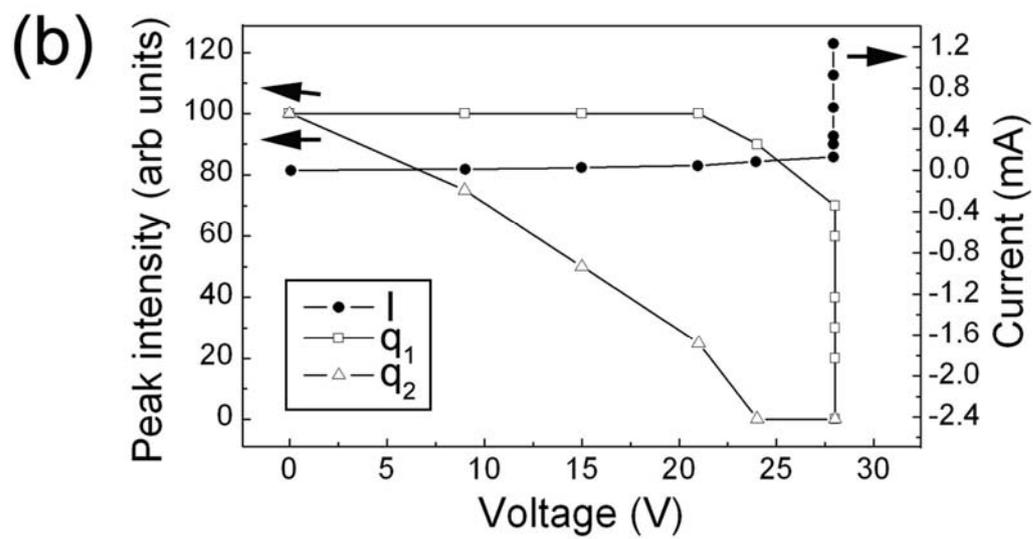